\documentstyle[12pt]{article}

\newcommand\res{{\rm Res}}

\begin{document}

\title{
ANTIBRACKET AS THE HAMILTONIAN STRUCTURE \\
OF A CLASSICAL INTEGRABLE SYSTEM}

\author{
Debashis Ghoshal\thanks{E-mail: ghoshal@mri.ernet.in}
\and Sudhakar Panda\thanks{E-mail: panda@mri.ernet.in}\\
{ }\\
{\it Mehta Research Institute} \\
{\it of Mathematics \&\ Mathematical Physics}\\
{\it 10 Kasturba Gandhi Marg}\\
{\it Allahabad 211 002, India}
}
\date{October 28,1994}
\maketitle

\centerline{Preprint MRI-PHY/20/94}

\begin{quote}
{\bf Abstract:} The time evolution in a supersymmetric extension of the
Kodomtsev-Petviashvilli hierarchy, a classical integrable system, is
shown to be Hamiltonian. The canonical bracket associated to the
Hamiltonian evolution is the classical analog of the antibracket
encountered in the quantization of gauge theories. This provides a new
understanding of supersymmetric Hamiltonian systems.
\end{quote}
\vfill

\newpage

Integrable models are an important class of dynamical system. The
Korteweg-de Vries (KdV) equation describing the motion of shallow water
waves, is one of an infinite hierarchy of nonlinear evolution equations
which are integrable\cite{dikii,das}. This hierarchy, which is completely
characterized by a second order differential operator, called a Lax
operator; has an infinite number of conserved quantities in common.
Furthermore the evolution equations can be written in two independent
canonical Hamiltonian form. Associated with these are two Poisson brackets
defining the so called bi-Hamiltonian structure.  The KdV hierarchy in
turn is generalized by considering higher order Lax operators. All these
integrable hierarchies are succintly described by the
Kodomtsev-Petviashvilli (KP) hierarchy based on a first order
pseudo-differential Lax operator.  Complete solution of the generalized
KdV hierarchy may be obtained from that of the KP hierarchy by imposing
suitable restrictions (see Ref.\cite{dikii} and references therein).

Integrable models can often be supersymmetrized. The supersymmetric
extension of the KP hierarchy was defined by Manin and
Radul\cite{mrskp}.  This system has an infinite number of conserved
Hamiltonians. In this letter we show that the evolution with respect to
bosonic time variables of the super KP hierarchy has a Hamiltonian
structure. (For an earlier attempt, see Ref.\cite{paro}.) The
associated bracket, however, is not the usual Poisson bracket, or more
precisely its graded generalization to include fermions\cite{htbook}.
Rather it is the classical analog of the {\it antibracket} proposed by
Batalin and Vilkovisky in their investigation of quantization of gauge
theories\cite{bv}.

An {\it antibracket} differs from a graded Poisson bracket in its
symmetry properties. It is symmetric between a pair of bosons, while
for a pair of fermions it is antisymmetric. We should recall here that
a graded Poisson bracket is (anti-)symmetric between a pair of fermions
(bosons respectively).  This difference is due to the fact that the
antibracket itself is fermionic.

The Batalin-Vilkovisky approach to gauge theories has greatly advanced our
conceptual understanding of them. (For more details and an extensive set
of references, see Ref.\cite{htbook}.) Now their appearance in the
canonical Hamiltonian equations of a classical integrable system allows us
to generalize the notion of Hamiltonian structure.  Although in the
following, we will focus only on the super KP hierarchy, this
feature will occur in other supersymmetric systems, for example
all integrable models obtained from an odd reduction of the
super KP hierarchy.

Let $x$ be a real variable and $\theta$ its supersymmetric partner. We
will collectively denote them by $z=(x,\theta)$. The
superderivative operator $D={\partial\over\partial\theta} +
\theta{\partial\over\partial x}$ satisfies the property
$D^2=\partial/\partial x$. The supersymmetric extension of the
KP hierarchy defined by Manin and Radul\cite{mrskp}
starts with the super-Lax operator
\begin{equation}
L = D + \sum_{i=0}^\infty u_{-i} D^{-i}, \label{laxop}
\end{equation}
where, $D^{-1}=\theta + \partial_x^{-1}{\partial\over\partial\theta}$
is the formal
inverse of $D$, that is, $D^{-1}D=DD^{-1}=1$. The Lax equations
\begin{equation}
{\partial L\over\partial t_{2n}} = [(L^{2n})_-,L] = [L,(L^{2n})_+],
\qquad n=1,2,\cdots,\label{laxeq}
\end{equation}
define an infinite number of non-linear evolution equations, with respect
to the (bosonic) `time' $t_{2n}$, ($n=1,2,\cdots$), for the superfields
$u_{-i}, i=0,1,\cdots$, by comparing the coefficients of $D^{-i}$.  The
evolution equations are consistent if and only if the constraint $(Du_0) +
2u_{-1}=0$ is satisfied\cite{mrskp}. Time $t_2$ is easily identified to
$-x$ from the observation that $(L^2)_+=D^2=\partial/\partial x$. The
super KP equation, after setting the fermionic variables to
zero, reduces to the original KP equation\cite{mrskp} describing
non-linear plasma waves. The equations of the KP hierarchy are well
known to admit soliton solutions\cite{dikii}.

The Lax operator $L$ is Grassmann odd. One can therefore introduce
a Grassmann parity for the fields $u_{-i}$ as $|u_{-i}|=i+1$ mod 2.
Therefore $u_{-i}$ with odd (even) superscript are bosonic (fermionic)
superfields respectively.

The hierarchy admits an infinite number of conserved quantities
(Hamiltonians) defined by
\begin{equation}
H_n = \int d^2z\, {\cal H}_n = {1\over n}\int d^2z\, \res
L^n,\qquad n=1,2,\cdots, \label{conserv}
\end{equation}
where, $\res L^n$ denotes the coefficient $A_{-1}$ of the $D^{-1}$ term in
the expansion of $L^n=\sum_{-\infty}^n A_{i} D^i$. For even integers
$n=2m$, $L^{2m} ={1\over 2} [L,L^{2m-1}]$ however, is an anticommutator
for which $\res$ is a total derivative\cite{mrskp} implying that $H_{2m}$
vanish for all $m\ge 1$. We therefore obtain non-trivial Hamiltonians
$H_{2m+1}$, $m\ge 0$, for $n=2m+1$. Whenever applicable, we will impose
the constraint $(Du_0) + 2u_{-1}=0$, to simplify the expressions.

The equation of motion for the field $u_i$, following from the Lax
equation (\ref{laxeq}) in its second form, can be written as a sum of
differential operators acting on the coefficients of $(L^{2n})_+$.
Using equation (\ref{conserv}), it is
easy to express these coefficients in terms of the Hamiltonians:
\begin{eqnarray}
L^{2n} &\equiv& \sum_{j=-\infty}^{2n} D^j p_j\nonumber\\
p_j &=& (-1)^j {\delta{\cal H}_{2n+1}\over\delta u_{-j-1}},\qquad
j=-1,0,1,\cdots,2n.\label{pjs}
\end{eqnarray}
In the above, the variational derivatives are defined by\cite{mrskp}
\begin{equation}
{\delta{\cal H}_{2n+1}\over\delta u_{-j}} = \sum_{k=0}^{\infty}
(-1)^{(j+1)k + {1\over 2}k(k+1)} \left(D^k\,{\partial{\cal H}_{2n+1}\over
\partial u_{-j}^{[k]}}\right), \label{variation}
\end{equation}
where $u^{[k]}=(D^k u)$. Notice that we can only determine the
coefficients $p_j$ for $j=-1,0,\cdots,2n$ this way. Fortunately, the Lax
equation (\ref{laxeq}) involves only these.

The evolution of the fields $u_{-i}$ can now be written in the
Hamiltonian form
\begin{equation}
{\partial u_{-i}\over\partial t_{2n}}(z) = \sum_{j=1}^\infty\int
d^2z'(-1)^{i+1}\{ u_{-i}(z), u_{-j}(z')\} {\delta{\cal
H}_{2n+1}\over\delta u_{-j}}(z'), \label{hamil}
\end{equation}
where we have denoted the differential operator by
$(-1)^{i+1}\{u_{-i},u_{-j}\}$, in anticipation of the fact that it leads to
a canonical bracket.

The equation (\ref{hamil}) does not involve the field $u_0$. We can,
however, add and subtract the $j=0$ term there and rewrite as
\begin{equation}
{\partial u_{-i}\over\partial t_{2n}}(z) = \{ u_{-i}(z),H_{2n+1}\}
+ (-)^i\int d^2z'\,\{u_{-i}(z),u_0(z')\}\,\res L^{2n}(z').
\label{hamilton}
\end{equation}
Here we have substituted the explicit form of $\delta {\cal
H}_{2n+1}/\delta u_0 = -p_{-1} = \res L^{2n}$ from Eq.(\ref{pjs}).

Comparing the Grassmann parity of both sides of the
Eq.(\ref{hamil}), we see that the bracket $\{f,g\}$ is
fermionic:
\begin{equation}
|\{f,g\}| = |f| + |g| + 1.\label{parity}
\end{equation}
Therefore the canonical bracket above cannot be a Poisson bracket. In fact,
Eq.(\ref{parity}) is the first property of an antibracket.

It is simple to calculate the canonical bracket between two fields,
following the standard procedure of Gelfand and Dikii\cite{dikii}. The
result is
\begin{eqnarray}
{\!\!\!\!\!\!\!}&{\!\!\!\!\!\!}&\{u_{-i}(z),u_{-j}(z')\} =
(-1)^{j+1}\sum_{k=0}^\infty (-1)^{k(i+j+1) + {1\over2}k(k+1)}
\times\nonumber\\
{\!\!\!\!\!\!\!}&{\!\!\!\!\!\!}&\left(\left[\!{k\!-\!j\atop k}\!\right]
D^k_z u_{k-i-j+1}(z) - (-)^{(i+1)(j+1)} \left[\!{k\!-\!i\atop k}\!\right]
D^k_{z'} u_{k-i-j+1}(z')\right)\Delta(z-z')
\label{antib}
\end{eqnarray}
where, $\Delta(z-z')=(\theta-\theta')\,\delta(x-x')$ is the delta function,
and the super-binomial coefficients $\left[{n\atop m}\right]$ have been
defined in Ref.\cite{mrskp}.

It is now straightforward to verify that the bracket above satisfies the
following symmetry property
\begin{equation}
\{f,g\} = - (-1)^{(|f|+1)(|g|+1)}\,\{g,f\}.\label{symm}
\end{equation}
In contrast to the Poisson bracket, the one above is (anti-)symmetric
between is pair of bosons (fermions). Between a boson and a fermion,
however, it is antisymmetric.

Finally, one can also check that the Jacobi identity
\begin{equation}
(-1)^{(|f|+1)(|h|+1)}\{f,\{g,h\}\} + \hbox{\rm cyclic} = 0
\label{jacobi}
\end{equation}
holds. Eqs.(\ref{parity},\ref{symm}) and (\ref{jacobi}) are all the
properties that a Batalin-Vilkovisky antibracket
satisfies\cite{bv,htbook}. The canonical bracket derived from
Eq.(\ref{hamil}) is therefore an antibracket.

In summary, we have shown that the evolution equations of the fields in
the super KP hierarchy defined by Manin and Radul, can be expressed in a
canonical Hamiltonian form. The canonical bracket associated to these
nonlinear equations of motion are the classical analog of the
Batalin-Vilkovisky {\it antibrackets}. This is unlike the Hamiltonian
systems known so far, where (a graded generalization of) a Poisson bracket
appears in the equation of motion.

An antibracket is known to define a Grassmann odd symplectic structure
in a phase space involving anticommuting variables\cite{schwarz}. The
canonical equations we have derived therefore also have a natural
geometrical interpretation. This generalizes the notion of classical
Hamiltonian formalism for systems involving anticommuting fields.


\begin{thebibliography}{9}
\bibitem{dikii}L. Dickey, {\it Soliton equations and Hamiltonian
systems}, Adv. Series in Math. Phys., vol 12, World Scientific (1991).
\bibitem{das}A. Das, {\it Integrable models}, World Scientific (1989).
\bibitem{mrskp}Y. Manin and A. Radul, {\it Commun. Math. Phys.}
{\bf 98}, 65 (1985).
\bibitem{paro}S. Panda and S. Roy, {\it Phys. Lett.} {\bf B291}, 77
(1992).
\bibitem{htbook}M. Henneaux and C. Teitelboim, {\it Quantization of gauge
systems}, Princeton University Press (1992).
\bibitem{bv}I.A. Batalin and G.A. Vilkovisky, {\it Phys. Rev.} {\bf
D28}, 2567 (1983), {\bf D30}, 508 (1984); {\it Nucl. Phys.} {\bf B234},
106 (1984).
\bibitem{schwarz}A. Schwarz, {\it Commun. Math. Phys.} {\bf 155}, 249
(1993), {\bf 158}, 373 (1993).
\end{thebibliography}
\end{document}